\documentclass[letterpaper,twocolumn,10pt]{article}
\usepackage{usenix,epsfig}
\usepackage{amssymb,comment,graphicx,hyperref,cite,array,mdwmath,mdwtab,fixltx2e}
\usepackage[normalem]{ulem}
\usepackage[cmex10]{amsmath}
\usepackage{algorithm,algorithmicx}
\newtheorem{definition}{Definition}

\newtheorem{proposition}{Proposition}

\RequirePackage[usenames,dvipsnames]{xcolor}
\newcommand{\changedcolor}{magenta}
\newcommand{\setchanged}{\color{\changedcolor}}
\newcommand{\changed}[1]{\texorpdfstring{{\setchanged{#1}}}{#1}}

\begin{document}

\title{\Large \bf Egalitarian computing\\(MTP 1.2)\\
1 Jan 2018}

\author{
{\rm Alex Biryukov}\\
University of Luxembourg\\alex.biryukov@uni.lu
\and
{\rm Dmitry Khovratovich}\\
University of Luxembourg\\khovratovich@gmail.com
} 
\maketitle

%

\subsection*{Abstract}
In this paper\footnote{This is a revised version of a paper published at USENIX Security Symposium in August 2016.} 
we explore several contexts where an adversary has an upper hand over the defender by using special hardware in an attack. These include password processing, hard-drive protection, cryptocurrency mining, resource sharing, code obfuscation, etc.

We suggest memory-hard computing as a generic paradigm, where
every task is amalgamated with a certain procedure requiring intensive access to RAM both in terms of size and (very importantly) bandwidth, so that transferring the computation to GPU, FPGA, and even ASIC brings little or no cost reduction. Cryptographic schemes that run in this framework become \emph{egalitarian} in the sense that both users and attackers are equal in the price-performance ratio conditions.

Based on existing schemes like {Argon2} and the recent generalized-birthday proof-of-work, we suggest a generic framework and two new schemes:
\begin{itemize}
\item \textsf{MTP}, a memory-hard Proof-of-Work based on the memory-hard function with fast verification and short proofs. It can be also used for memory-hard time-lock puzzles. The MTP-Argon2 requires 2 GB of RAM to make a proof, and can be initialized and start producing proofs in less than 1 second -- performance hardly beaten by competitors like Equihash for the same amount of RAM.
\item \textsf{MHE}, the concept of memory-hard encryption, which utilizes  available RAM to strengthen the encryption for the low-entropy keys (allowing to bring back 6 letter passwords).
\end{itemize}

The MTP PoW has been tweaked in October-November 2017 to withstand certain recently found attacks~\cite{DinurN17,Bevand17}. 

\textbf{Keywords}: MTP, MHE, Argon2, memory-hard,  asymmetric, proof-of-work, botnets, encryption, time-lock puzzles.

\section{Introduction}\label{sec:intro}

\subsection{Motivation}

Historically attackers have had more resources than defenders, which is still mostly true. Whether it is secret key recovery or document forgery, the attackers are ready to spend tremendous amount of computing power to achieve the goal. In some settings it is possible to make most attacks infeasible by simply setting the key length to 128 bits and higher. In other settings the secret is limited and the best the defender can do is to increase the time needed for the attack, but not to render the attack impossible.

Passwords, typically stored in a hashed form, are a classical example. As people tend to choose passwords of very low entropy,  the security designers added unique salts and then increased the number of hash iterations. In response the attackers switched to dedicated hardware for password cracking, so that the price of single password recovery dropped dramatically, sometimes by a few orders of magnitude. 

A similar situation occurred in other contexts. The Bitcoin cryptocurrency relies on continuous preimage search for the SHA-256 hash function, which is much cheaper on custom ASICs, consuming up to 30,000 times~\footnote{We compare $2^{32}$ hashes per joule on the best ASICs with $2^{17}$ hashes per joule on the most efficient x86-laptops} less energy per solution than most efficient x86 laptops~\cite{bitcoinhard}. Eventually, the original concept of an egalitarian cryptocurrency~\cite{nakamoto09} vanished with the emergence of huge and centralized mining pools.

Related problems include password-based key derivation for hard-drive encryption, where the data confidentiality directly depends on the password entropy, and where offline attack is exceptionally easy once the drive is stolen. Similar situation arise in the  resource sharing and spam countermeasures. In the latter it is  proposed that  every user presents a certain proof (often called proof-of-work), which should be too expensive for spammers to generate on a large scale. Yet another setting is that of code obfuscation, in which powerful reverse-engineering/de-compilation tools can be used in order to lift the proprietary code or secrets embedded in the software.

\subsection{Egalitarian computing}
Our idea is to remedy the disparity between ordinary users and adversaries/cheaters, where latter could use botnets, GPU, FPGA, ASICs to get an advantage and run a cheaper attack. We call it \emph{egalitarian computing} as it should establish the same price for a single computation unit on all platforms, so that the defender's hardware is optimal both for attack and defence. Equipped with egalitarian crypto schemes, defenders may hope to be on par with the most powerful attackers.

The key element of our approach is large (in size) and intensive (in bandwidth) use of RAM as a widely available and rather cheap unit for most defenders.  In turn, RAM is rather expensive on FPGA and ASIC\footnote{The memory effect on  ASICs can be illustrated as follows. A compact 50-nm DRAM implementation~\cite{giridhar2013dram} takes 500 mm$^2$ per GiB, which is equivalent to about 15000 10 MHz  SHA-256 cores in the best Bitcoin 40-nm ASICs~\cite{bitasic} and is comparable to a CPU size. Therefore, an algorithm requiring 1 GiB for 1 minute would have the same AT cost as an algorithm requiring $2^{42}$ hash function calls,
whereas the latter can not finish on a PC even in 1 day. In other words, the use of memory can increase the AT cost by a factor of 1000 and more at the same time-cost for the desktop user.}, and slow on GPU, at least compared to memoryless computing tasks. All our schemes  use a lot of memory and a lot of bandwidth --- almost as much as possible.

We suggest a single framework for this concept and concrete schemes with an unique combination of features. 
In the future, adoption of our concept could allow a homogenization of computing resources, a simplified security analysis, and relaxed security requirements. When all attackers use the same hardware as defenders, automated large-scale attacks are no longer possible. Shorter keys, shorter passwords, faster and more transparent schemes may come back to use. 

\paragraph{Related work}

The idea of extensive memory use in the context of spam countermeasures dates back at least to 2003~\cite{DworkGN03,AbadiBW03} and was later 
refined in~\cite{DworkNW05}. Fast memory-intensive hash functions were proposed first by Percival in 2009~\cite{percival2009stronger} and later among the submissions of the Password Hashing Competition. Memory-intensive proofs-of-work have been studied both in theory~\cite{DziembowskiFKP13} and practice~\cite{Andersen14,Tromp14}.

\paragraph{Paper structure} We describe the goals of our concept and give a high level overview in Section~\ref{sec:framework}. Then we describe existing applications where this approach is implicitly used: password hashing and cryptocurrency proofs of work (Section~\ref{sec:app}). We present our own progress-free Proof-of-Work MTP with fast verification, which can also serve as a memory-hard time-lock puzzle,  in Section~\ref{sec:mtp}.
The last Section~\ref{sec:mhe} is devoted to the novel concept of memory-hard encryption, where we present our scheme MHE aimed to increase the security of password-based disk encryption. Appendix~\ref{sec:AppB} describes the differences with the original MTP v1.0 paper.

\section{Egalitarian computing as  framework}\label{sec:framework}

\subsection{Goal}

Our goal is to alter a certain function $\mathcal{H}$ in order to maximize its computational cost on the most efficient architecture -- ASICs, while keeping
the running time on the native architecture (typically x86) the same. We ignore the design costs due to nontransparent prices, but instead estimate the running costs by measuring the time-area product~\cite{Thompson79,BernsteinL13}.  On ASICs the memory size $M$ translates into certain  area $A$. The ASIC running time $T$ is determined by the length of the longest computational chain and by the ASIC memory latency.

Suppose that an attacker wants to compute $\mathcal{H}$ using only a fraction $\alpha M$ of memory for some $\alpha<1$. Using some tradeoff specific to $\mathcal{H}$, he has to spend $C(\alpha)$ times as much computation and his running time increases by   the factor $D(\alpha)$ (here $C(\alpha)$ may exceed $D(\alpha)$ as the attacker can parallelize the computation). In order to fit the increased computation into time, the attacker has to place $\frac{C(\alpha)}{D(\alpha)}$ additional cores on chip. Therefore, the  time-area product changes from $ AT_{1}$ to $AT_{\alpha}$ as 
 \begin{multline}\label{eq:at}
 AT_{\alpha} =A\cdot\Big(\alpha+ \frac{\beta C(\alpha)}{D(\alpha)}\Big)T\cdot D(\alpha) =\\ = AT_1(\alpha D(\alpha) + C(\alpha)\beta),
  \end{multline}
where $\beta$ is the fraction of the original memory occupied by a single computing core. If the tradeoff requires significant communication between the computing cores, the memory bandwidth limit $Bw_{max}$ may also increase the running time. In practice we will have $D(\alpha) \geq  C(\alpha)\cdot Bw/Bw_{max}$,
where $Bw$ is the  bandwidth for $\alpha=1$.

\begin{definition} 
We call function $\mathcal{F}$  \emph{memory-hard} (w.r.t. $M$) if any algorithm $\mathcal{A}$ that computes $\mathcal{H}$ using $\alpha M$ memory has the computation-space tradeoff  $C(\alpha)$ where $C()$ is at least a superlinear function of $1/\alpha$.
\end{definition}
 It is known~\cite{HopcroftPV77} that any function whose computation is interpreted as a directed acyclic graph with $T$ vertices of constant in-degree, can be computed using $O(\frac{T}{\log T})$ space, where the constant in $O()$ depends on the degree. However, for concrete hash functions very few tradeoff strategies have been published, for example~\cite{trade-att}.

\subsection{Framework}

Our idea is to combine a certain computation $\mathcal{H}$ with a memory-hard function $\mathcal{F}$. This can be done by modifying $\mathcal{H}$ using input from $\mathcal{F}$ (\emph{amalgamation}) or by transforming its code to an equivalent one (\emph{obfuscation}).

The \textbf{amalgamation} is used as follows. The execution of $\mathcal{H}$ is typically a sequence of smaller steps $H_i, i<T$, which take the output $V_{i-1}$ from the previous step and produce the next output $V_i$. For our purpose we need another primitive, a memory-hard function $\mathcal{F}$, which fills the memory with some blocks $X[i], i<T$. We suggest combining $\mathcal{H}$ with $\mathcal{F}$,
for example like:
$$
\mathcal{H}' =  H_T'\circ H_{T-1}'\circ \cdots \circ H_1',
$$
where
$$
H_i'(V_{i-1}) = H(V_{i-1} \oplus X[i-1]).
$$
Depending on the application, we may also modify $X[i]$ as a function of $V_{i-1}$ so that it is impossible to precompute $\mathcal{F}$. The idea is that any computation of $\mathcal{H}'$ should use $T$ blocks of memory, and if someone wants to use less, the memory-hardness property would impose computational penalties on him. This approach will also work well for any code that uses nonces or randomness produced by PRNG. PRNG could then be replaced by (or intermixed with the output of) $F$.

The \textbf{obfuscation} principle works as follows. Consider a compiler producing an assembly code for some function $\mathcal{H}$. We make it to run a memory-hard function  $\mathcal{F}$ on
a user-supplied input $I$ (password)
and produce certain number of memory blocks. For each if-statement of the form

\begin{tabular}{cc}
\textbf{if} $x$ &\textbf{then} A\\
&\textbf{else} B\\
\end{tabular}

the compiler computes a \emph{memory-hard bit} $b_i$ which is extracted from the block $X[i]$ (the index can also depend on the statement for randomization) and alters the statement as

\begin{tabular}{cc}
\textbf{if} $x\oplus b_i$ &\textbf{then} A\\
&\textbf{else} B\\
\end{tabular}

for $b_i=0$ and

\begin{tabular}{cc}
\textbf{if} $x\oplus b_i$ &\textbf{then} B\\
&\textbf{else} A\\
\end{tabular}

for $b_i=1$. This guarantees that the program will have to run $\mathcal{F}$ at least once (the bits $b_i$ can be cached if this if-statement is used multiple times, ex. in a loop).

Accessing the memory block from a random memory location for each conditional statement in practice would slow down the program too much, so compiler can perform a 
tradeoff depending on the length of the program, the number of conditional statements in it and according to a tunable degree of required
memory-hardness for a program. Memory-hard bits could be mixed into  opaque predicates or other code obfuscation constructs like code-flattening logic.

We note that in order for a program to run correctly, the user needs to supply correct password for  $\mathcal{F}$, even though the source code of the program is public. A smart decompiler, however,  when supplied with the password, can obtain clean version of the program by running $\mathcal{F}$ only once. 


Our schemes described in the further text use the amalgamation principle only, so we leave the research directions in obfuscation for future work.

\section{Egalitarian computing in applications}\label{sec:app}

In this section we outline several applications, where memory-hard functions are actively used in order to achieve egalitarian computing.

\subsection{Password hashing with a memory-hard function}\label{sec:ph}

The typical setting for the password hashing is as follows. A user selects a password $P$ and submit it to the authentication server with his identifier $\mathrm{Id}$. The server hashes $P$ and unique salt
$S$ with some function $\mathcal{F}$, and stores $(\mathrm{Id},S,\mathcal{F}(P,S))$ in the password file. The common threat is the password file theft, so that an attacker can try the passwords from his 
dictionary file $D$ and check if any of them yields the stolen hash. The unique $S$ ensures that the hashes are tried one-by-one.

Following massive password cracking attacks that use special hardware~\cite{fpga,SprengerB12}, the security community initiated the Password Hashing Competition~\cite{comp} to select the hash  function that withstands the most powerful adversaries. The Argon2 hash function~\cite{Argon2} has been recently selected as the winner. We stress that the use of memory-hard function for password hashing does not make the dictionary attacks infeasible, but it makes them much more expensive in terms of the single trial cost.

\paragraph{Definition and properties of Argon2} We use  Argon2 in our new schemes, which are described in Sections~\ref{sec:mtp} and \ref{sec:mhe}. The original Argon2 design has been modified (with all differences explicitly mentioned below) to withstand more powerful adversaries in the Proof-of-work setting. Here we outline the key  elements of the Argon2 design
that are used in our scheme. For more details and their rationale we refer the reader to~\cite{Argon2}.

Argon2 takes $P$ (password), $S$ (salt), and possibly some additional data $U$ as inputs. It is parametrized by the memory size $T$,  number of iterations  $t$, and the available parallelism $l$. It fills 
$T$ blocks of memory $X[1],X[2], \ldots,X[T]$ (1 KiB each) and then overwrites them $(t-1)$ times. Each block $X[i]$ is generated using internal compression function $F$, which takes $X[i-1]$, 
 $X[\phi(i)]$. For $t=1$ this works as follows, where $H$ is the 256-bit Blake2b, $H_{2048}$ is a 2048-bit  hash function based on $H$.
\begin{equation}
\label{eq:dag}
\begin{aligned}
H_0 &\leftarrow H(P,S,U);\\
X[0]||X[1] &\leftarrow H_{2048}(H_0);\\
X[i] &\leftarrow F(X[i-1],X[\phi(i)]),\; i>1;\\
Out&\rightarrow H(X[T-1]).
\end{aligned}
\end{equation}
The indexing function $\phi(i)$ is defined separately for each of two versions of Argon2: 2d and 2i. The Argon2d, which we use in this paper, computes it as a function of the previous block $X[i-1]$.

The authors proved~\cite{Argon2} that all the blocks are generated distinct assuming certain collision-resistant-like properties of $F$. They also reported the performance of 0.7 cpb on the Haswell CPU with 4 threads, and 1.6 cpb with 1 thread.

\paragraph{Multi-lane Argon2} Argon2  supports multithreading by arranging its blocks into multiple $p$ lanes and constructing the $\phi$ function so that only not so recent blocks are referenced. The blocks are indexed in a two-dimensional array with $p$ rows and $T/p$ columns. For non-ambiguity, we suggest the following mapping between one- and two-dimensional representations:
\begin{align}
\label{eq:blockindex}
    \psi: [0;T-1] &\rightarrow [0;p-1]\times [0;T/p-1];\\
    \psi (i) &= (p\cdot i/T,i\bmod{(T/p)});\\
    \psi^{-1}(j,k) &= jT/p+k.
\end{align}

\paragraph{Tradeoff security of Argon2} Using the tradeoff algorithm published in~\cite{trade-att}, the authors report the  values  $C(\alpha)$ (average computational tree size) and $D(\alpha)$ (average tree depth) up to $\alpha=1/7$ with $t=1$. It appears that $C(\alpha)$ is exponential in $\alpha$, whereas $D(\alpha)$ is linear.

\begin{table}[hb]
\renewcommand{\arraystretch}{1.3}
$$
\begin{array}{|c||c|c|c|c|c|c|}
\hline
\text{$\alpha$ } &\frac{1}{2} &\frac{1}{3} &\frac{1}{4} &\frac{1}{5} &\frac{1}{6} &\frac{1}{7}  \\
\hline
\text{$C(\alpha)$} &1.5& 4& 20.2& 344&  4660 &  2^{18}\\
\text{$D(\alpha)$} & 1.5 & 2.8 & 5.5 & 10.3 & 17 &27 \\
\hline
\end{array}
$$
\caption{Time and computation penalties for the ranking tradeoff attack for  Argon2d.}\label{tab:generic3}
\end{table}

\subsection{Proofs of work}

A \emph{proof-of-work scheme}  is a challenge-response  protocol,
where one party (Prover) has to prove (maybe probabilistically) that it has performed a certain amount of computation following a request from another party (Verifier).
It typically relies on  a  computational problem where a solution is assumed to have fixed cost, such as the preimage search    in the Bitcoin protocol and other cryptocurrencies. Other applications may include spam protection, where a proof-of-work is a certificate that is easy to produce for ordinary sender, but hard to generate in large quantities given a botnet (or more sophisticated platform).

The proof-of-work algorithm must have a few properties to be suitable for cryptocurrencies:
\begin{itemize}
\item It must be \emph{amortization-free}, i.e. producing $q$ outputs for $\mathcal{B}$ should be $q$ times as expensive;
\item The solution must  be \emph{short} enough and verified quickly using \emph{little memory} in order to prevent DoS attacks on the verifier.
\item The time-space tradeoffs must be \emph{steep} to prevent any price-performance reduction.
\item  The time and memory parameters must be \emph{tunable independently} to sustain constant mining rate. 
\item  To avoid a clever prover getting advantage over the others the advertised algorithm must be the most efficient algorithm to date (\emph{optimization-freeness}).
\item The algorithm must be \emph{progress-free} to prevent centralization: the mining process must be stochastic so that the probability to find a solution grows steadily with time and is non-zero for small time periods. 
\item Parallelized implementations must be limited by the memory bandwidth.
\end{itemize}

 As demonstrated in~\cite{Holiday}, almost any hard problem can be turned into a proof-of-work, even though it is difficult to fulfill all these properties. The well-known hard and NP-complete problems   are  natural candidates, since the best algorithms for them run in (sub)exponential time, whereas the verification is polynomial. The   proof-of-work scheme Equihash~\cite{Holiday} is built on  the \emph{generalized-birthday}, or $k$-XOR, problem, which looks for a set of $n$-bit strings that XOR to zero. The best existing algorithm is due  to Wagner~\cite{Wagner02}.  This problem is particularly interesting, as  the time-space tradeoff steepness can be adjusted by changing $k$, which does not hold, e.g., in  hard knapsacks. 

\paragraph{Drawbacks of existing PoW}
We briefly discuss existing alternatives here. The first PoW schemes by Dwork and Naor~\cite{DworkN92} were just computational problems with fast verification such as the square root computation, which do not require large memory explicitly. The simplest scheme of this kind is Hashcash~\cite{back2002hashcash}, where a partial preimage to a cryptographic hash function is found (the so called \emph{difficulty test}). Large memory comes into play in~\cite{DworkGN03}, where a random array is shared between the prover and the verifier thus allowing only large-memory verifiers. This condition was relaxed in~\cite{DworkNW05}, where superconcentrators~\cite{Pippenger77} are used to generate the array, but the verifier must still hold large memory in the initialization phase. Superconcentrators were later used in the Proof-of-Space construction~\cite{DziembowskiFKP13}, which allows fast verification. However, the scheme~\cite{DziembowskiFKP13} if combined with the difficulty test is vulnerable to cheating (see Section~\ref{sec:simple} for more details) and thus can not be converted to a progress-free PoW. We note that the superconcentrators make both ~\cite{DworkNW05} and ~\cite{DziembowskiFKP13} very slow.

Ad-hoc  but faster schemes started with scrypt~\cite{percival2009stronger}, but fast verification is possible only with rather low amount of memory. Using more memory (say, using Argon2~\cite{Argon2}) with a difficulty test but verifying only a subset of memory is prone to cheating as well (Section~\ref{sec:simple}).  

The scheme~\cite{Holiday} is quite promising and is now employed in Zcash~\cite{Zcash}. The fastest implementation reported needs 0.5 second to get a proof that certifies the memory allocation of 150 MiB. The performance for higher memory requirements is expected to grow at least linearly. As a result, the algorithm is not truly progress-free: the probability that the solution for a-few-GiB-memory-challenge is found within the first few seconds is actually zero. It can be argued that this could stimulate centralization among the miners.

Ethash, the PoW currently employed in Ethereum~\cite{EthereumYellow}, operates similarly to~\cite{DworkGN03,DworkNW05}: it generates a pseudo-random array once every few days, and then makes a random walk over the entries until a certain property is satisfied. The proofs are short but the initialization stage takes several minutes nowadays.  This makes the Ethereum mining process more centralized as a miner would have to skip a few blocks after he joins. In addition, the ad-hoc nature of Ethash makes it difficult to claim its memory-hardness. As the Ethash is (before the inception of Casper) a widely deployed PoW, we summarize its features and drawbacks here:
\begin{itemize}
    \item It requires about 4 GB of RAM in the read-only manner (ROM): once filled, it remains untouched for days. Thus it is suitable for mining on platforms with low write speed as writing is seldom required;
    \item Ethash  is an ad-hoc design, there is no proof (or even sketch) that it can not be handled with less memory.
    \item When Ethash is used inside a multi-currency miner, it might take significant time to switch between the currencies causing profit loss.
    \item The verification is not totaly memoryless: a volume of about 32 MiB is needed to verify the PoW. 
\end{itemize}
Additional analysis of Ethash (somewhat outdated by now) can be found in~\cite{EthashAnalysis}.

Finally, we mention schemes Momentum~\cite{momentum} and Cuckoo cycle~\cite{Tromp14}, which provide fast verification due to their combinatorial nature. They rely on the memory requirements for the collision search  (Momentum) or graph cycle finding (Cuckoo). However, Momentum is vulnerable to a sublinear time-space tradeoff~\cite{Holiday}, whereas the first version of the Cuckoo scheme was broken in~\cite{Andersen14}.

We summarize the properties of  the existing proof-of-work constructions in Table~\ref{tab:pow}. The AT cost is estimated for the parameters that enable 1-second generation time on a PC.
\begin{table*}
\begin{center}
\renewcommand{\arraystretch}{1.3}
\begin{tabular}{|c||c|c|c|c|c|c|c|}
\hline
{Scheme} &AT cost &Speed &\multicolumn{2}{c}{Verification}&Tradeoff &Paral-sm& Progress\\
& & & Fast & M/less & &&-free\\
\hline
Dwork-Naor I~\cite{DworkN92} & Low& High & Yes  & Yes  & Memoryless &  Yes  & Yes\\
Dwork-Naor II~\cite{DworkGN03} & High& Low & Yes  & No   & Memoryless & Constr. & Yes \\
Dwork-Naor III~\cite{DworkNW05} & Medium& Low & Yes  & No   & Exponential & Constr. & Yes  \\
Hashcash/Bitcoin~\cite{back2002hashcash} & Low & High & Yes & Yes & Memoryless & Yes & Yes\\
Pr.-of-Space~\cite{DziembowskiFKP13}+Diff.test & High & Low & Yes & Yes & Exponential & No & No\\
Litecoin  & Medium & High & Yes & Yes & Linear & No & Yes\\
Argon2-1GiB + Diff.test & High & High & No & No & Exponential & No & Yes\\
Momentum~\cite{momentum} &Medium& High& Yes &Yes & Attack~\cite{OorschotW99,Holiday} & Yes & Yes\\
Cuckoo cycle~\cite{Tromp14} & Medium~\cite{Andersen14} & Medium & Yes & Yes & Linear~\cite{Andersen14} & Yes & Partly\\
Equihash~\cite{Holiday} & High & Medium & Yes & Yes & Exponential & Constr. & Yes\\
Ethash~\cite{EthereumYellow} & High & Low & Yes & Partly & Exponential & Constr. & Yes\\
\hline
MTP & High & High & Yes & Yes & Exponential & Constr. & Yes\\
\hline
\end{tabular}\\[10pt]\caption{Review of existing proofs of work. Litecoin utilizes  scrypt with 128KiB of RAM followed by the difficulty test). M/less -- memoryless; constr. -- constrained.}\label{tab:pow}
\end{center}
\end{table*}

\section{MTP: Proofs of work and time-lock puzzles based on memory-hard function}\label{sec:mtp}

In this section we present a novel proof-of-work algorithm MTP (for Merkle Tree Proof) with
fast verification, which in particular solves the progress-free problem of~\cite{Holiday}. Our approach is based on the
memory-hard function, and the concrete proposal involves Argon2.  

Since fast memory-hard functions $\mathcal{F}$ such as Argon2 perform a lengthy chain of computations, but do not solve any NP-like problem, it is not fast to verify that $Y$ is the output of $F$. Checking some specific (say, last) blocks does not help, as explained in detail in the further text. We thus have to design a scheme that lower bounds the time-area product for the attacker, even if he computes a slightly modified function.

\subsection{Description of MTP}

\begin{figure}[ht]
  \ifpdf
\begin{center}
  \includegraphics[scale=0.4]{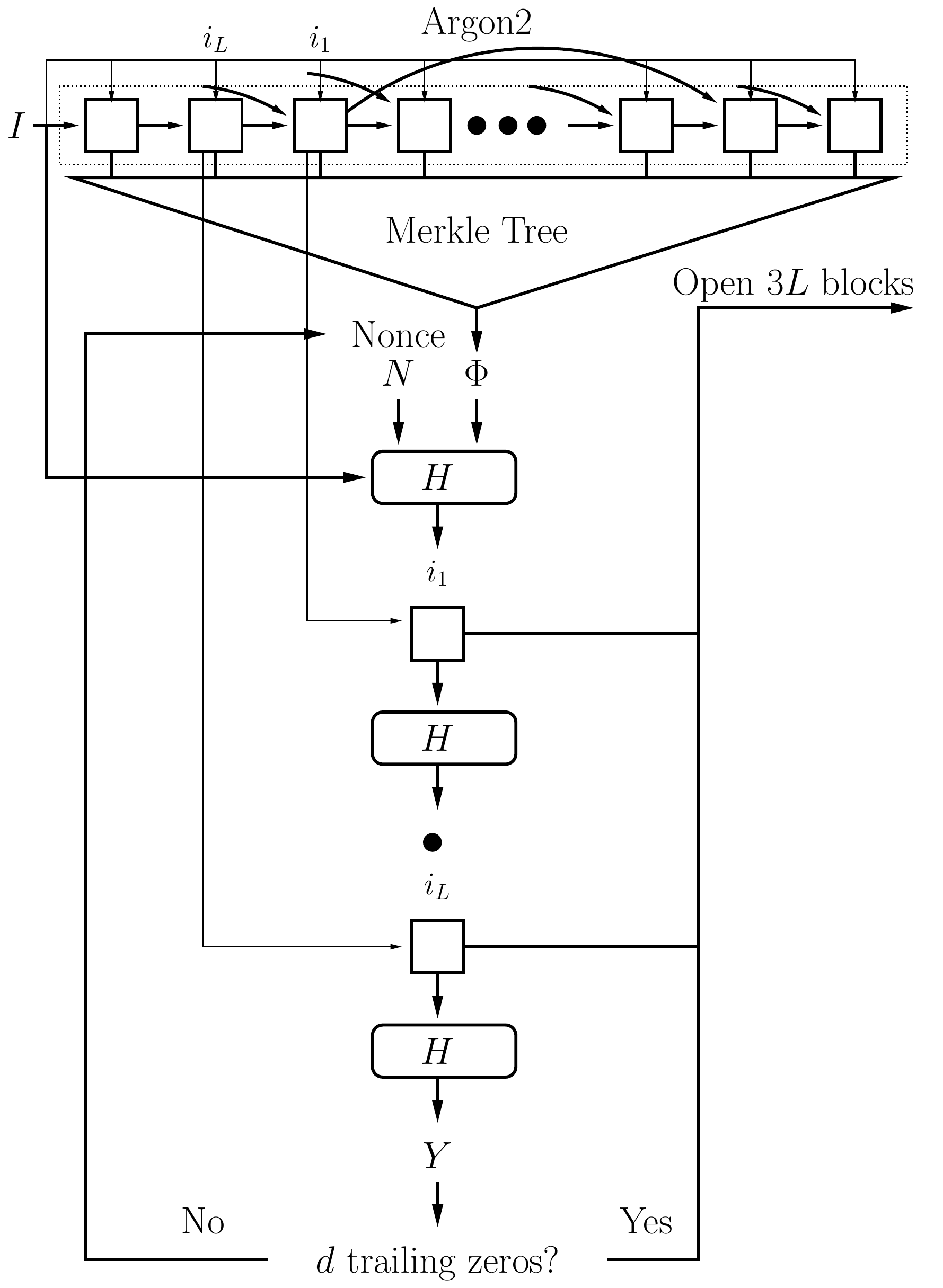}
  \caption{MTP: Merkle-tree based Proof-of-Work with light verification. }\label{fig:instant}
\end{center}
\fi
  \end{figure}

 Consider a  memory-hard function $\mathcal{F}$ that satisfies Equation~\eqref{eq:dag} (for instance, Argon2) with a single pass over the memory producing $T$ blocks with the internal compression function $F_I(X,Y)$ (parameterized by the challenge $I$) and a
 cryptographic hash function $H$.
 We propose the following non-interactive protocol for the Prover (Figure~\ref{fig:instant}) in Algorithm~\ref{alg:mtp}, where $L$ and $d$ are security parameters.  The average number of calls to $F$ is $T+2^dL$.
 
\begin{algorithm}
\textbf{Input:} Challenge $I$, parameters $L,d$.
\begin{enumerate}
 \item Compute $\mathcal{F}(I)$ using $F_I$ and store its $T$ blocks $X[0]$, $X[1]$, $\ldots$, $X[T-1]$ in the memory. 
\item Compute the root $\Phi$ of the Merkle hash tree  (see Appendix~\ref{sec:tree}).
\item Select nonce $N$.
\item Compute  $Y_0 = H(I,\Phi,N)$ where $H$ is a cryptographic hash function.
\item For $1 \leq j \leq L$:
$$
\begin{aligned}
i_j &\leftarrow Y_{j-1} \pmod{T};\\
Y_j &\leftarrow H(Y_{j-1},X[i_j]).
\end{aligned}
$$
\item If $Y_L$ has $d$ trailing zeros, then $(I,\Phi,N,\mathcal{Z})$ is the proof-of-work, where $\mathcal{Z}$ consists of $3L$  openings of  blocks $\{X[i_j-1],X[\phi(i_j)], X[i_j]\}_j$ together with their indices, where the block $X[i_j]$ itself is not included, only its opening.  Otherwise go to Step 3.
\end{enumerate}
\textbf{Output:} Proof $(I,\Phi,N,\mathcal{Z})$.
\caption{MTP: Merkle-tree based Proof-of-Work. Prover's algorithm}\label{alg:mtp}
\end{algorithm}

 The verifier, equipped with $\mathcal{F}$ (and $F$ as part of its description) and $H$, runs Algorithm~\ref{alg:mtpv}.
 
\begin{algorithm}
\textbf{Input:} Proof $(I,\Phi,N,\mathcal{Z})$, parameters $L,d$.
\begin{enumerate}
 \item Compute  $Y_0 = H(I,\Phi,N)$.
 \item For $\mathcal{Z}$ for $1 \leq j \leq L$ :
\begin{itemize}
    \item Compute $i_j  \leftarrow Y_{j-1} \pmod{T}$;
    \item Retrieve $X[i_j-1],X[\phi(i_j)]$ from the proof;
    \item Compute $ X[i_j] = F_I(X[i_j-1],X[\phi(i_j)])$;
    \item Verify openings for $X[i_j-1],X[\phi(i_j)], X[i_j]$ and their respective
    positions $i_j-1,\phi(i_j),i_j$, where $\phi(i_j)$ is computed out of $X[i_j-1]$;
    \item  Compute $Y_j \leftarrow H(Y_{j-1},X[i_j])$.
\end{itemize}
\item Check whether  $Y_L$ has $t$ trailing zeros.
\end{enumerate}
\textbf{Output:} Yes/No.
\caption{MTP: Verifier's algorithm}\label{alg:mtpv}
\end{algorithm}
 
\subsection{Cheating strategies}\label{sec:cheat}

Let the computation-space tradeoff for $\mathcal{H}$ and the default memory value $T$ be given by functions $C(\alpha)$ and $D(\alpha)$ (Section~\ref{sec:framework}).

\paragraph{Memory savings} Suppose that a cheating prover wants to reduce the AT cost by using $\alpha T$ memory for some $\alpha <1$. First, he computes $\mathcal{F}(I)$ and $\Phi$, making $C(\alpha)T$ calls to $F$. Then for each $N$ he has to get or recompute $L$ blocks using only $\alpha T $ stored blocks. The complexity of this step is equal to the complexity of recomputing random $L$ blocks during the first computation of $\mathcal{F}$. A random  block is recomputed by a tree of average size $C(\alpha)$ and depth $D(\alpha)$. Therefore, to compute the proof-of-work, a memory-saving prover has to make $C(\alpha)(T+2^dL)$ calls to $F$, so his amount of work grows by a factor $C(\alpha
)$.

\paragraph{Block modification} The second cheating strategy is to compute a different function $\widetilde{\mathcal{F}}\neq \mathcal{F}$. More precisely, the cheater produces some blocks $X[i']$ (which we call \emph{inconsistent}  as in~\cite{DziembowskiFKP13}) not as specified by Equation~\eqref{eq:dag} (e.g. by simply computing $X[i'] = H(i')$). In contrast to the verifiable computation approach, our protocol allows a certain number of inconsistent blocks. Suppose that the number of inconsistent blocks is $\epsilon T$,
then the chance that no inconsistent block is detected by $L$ opened blocks is 
$$
\gamma=(1-\epsilon)^L.
$$
Therefore, the probability for a proof-of-work with $\epsilon T$ inconsistent blocks to pass the opening test is $\gamma$. In other words, the cheater's time is increased by the factor $1/\gamma$. We note that it does not make sense to alter the blocks after the Merkle tree computation, as any modified block would fail the opening test.

\paragraph{Skewed blocks} The third cheating strategy has been inspired by~\cite{DinurN17}. Inconsistent blocks split the entire chain of blocks into consistent chunks. An attacker tries several candidates for the inconsistent block (the first in the chunk) to make some next blocks in the chunk referring to inconsistent or stored blocks, so that he looks them up with no penalty in the last phase of the computation. Suppose that the fraction of such  additional blocks, which we call \emph{ skewed}, is $\delta$, so the average consistent chunk of length $1/\epsilon$ would have $\delta/\epsilon$ such blocks. As each block would require $\frac{1}{\alpha + \epsilon}$ attempts to make it refer to inconsistent or stored blocks,  each attempt costs at least 1 call to $F$, and these efforts multiply up for one chunk. Thus an attacker would make at least $\frac{1}{(\alpha + \epsilon)^{\delta/\epsilon}}$ extra calls to $F$ per chunk to make it work. 

\paragraph{Overall cheating penalties} Let us accumulate the three cheating strategies into one. Suppose that a cheater stores $\alpha T$ blocks,  additionally allows
$\epsilon T$ inconsistent blocks, and makes $\delta T$ other blocks to refer to either stored or inconsistent ones. His penalty while computing $T$ blocks will be at least  $C(\alpha+\epsilon+\delta)$ per block plus extra $(\alpha + \epsilon)^{-\delta/\epsilon}$ per chunk due to skewed blocks.  His penalty when looking up $L $ blocks would be at least $C(\alpha+\epsilon+\delta)$ per block plus on average extra $\frac{\delta}{2\epsilon}$ calls per skewed block. Thus he  makes at least
\begin{equation}\label{eq:incons}
\frac{\left(C(\alpha+\epsilon+\delta)+\frac{\epsilon}{(\alpha + \epsilon)^{\delta/\epsilon}}\right)T+2^dL(C(\alpha+\epsilon+\delta)+\frac{\delta^2}{2\epsilon})}{\gamma}
\end{equation}
calls to $F$.  The concrete values are determined by the penalty function $C()$, which depends on $\mathcal{F}$.  The depth penalty would be $D(\alpha+\epsilon+\delta)$ for $T$ blocks and $D(\alpha+\epsilon+\delta)+\frac{\delta^2}{2\epsilon}$ for the $L$ blocks.

\subsection{Parallelism}

Both honest prover and cheater can parallelize the computation for $2^t$ different nonces. However, the latency of cheater's computation will be higher, since each  block generates a recomputation tree of average depth $D(\alpha+\epsilon+\delta)$.

 \subsection{Why simpler approach does not work: grinding attack} \label{sec:simple}
 Now we can explain in more details why the composition of $\mathcal{F}$ and the difficulty test is not a good proof-of-work even if some internal blocks of $\mathcal{H}$ are opened.
 Suppose that the proof is accepted if $H(X[T])$ has certain number $d$ of trailing zeros. One would expect that a prover has to try  $2^d$ distinct $I$ on average and thus call $\mathcal{F}$ $2^d$ times to find a solution. However, a cheating prover can simply try $2^d$ values for $X[T]$ and find one that passes the test in just $2^d$ calls to $H$. Although $X[T]$ is now inconsistent, it is unlikely to be selected among $L$ blocks to open, so the cheater escapes detection easily. Additionally checking $X[T]$ would not resolve the problem since a cheater would then modify the previous block, or $X[\phi(T)]$, or an earlier block and then propagate the changes. A single inconsistent block is just too difficult to catch\footnote{We have not seen any formal treatment of this attack in the literature, but it appears to be known in the community. It is mentioned in~\cite{ParkPAFG15} and \cite{MillerCohen16}.}.

\subsection{Modified Argon2}
We now modify Argon2 to strengthen it against PoW-specific cheaters.
In the original  Argon2  the  compression function works as 
$$
X[i] \leftarrow F(X[i-1],X[\phi(i)]).
$$
We modify it so that it takes the parameters  $i$ (block index) and the hash of the original challenge $H_0$. The new compression function is denoted as
$$
X[i] \leftarrow F_{H_0,i}(X[i-1],X[\phi(i)]).
$$
and defined as follows (the difference to the original one are shown in \changed{\changedcolor}.
\begin{enumerate}
    \item Compute $R \leftarrow X[i-1] \oplus X[\phi(i)]$;
    \item Treat $R$ as a  $8\times 8$-matrix of 16-byte registers $R_0, R_1,\ldots, R_{63}$.
    \changed {Set $R_7 \leftarrow \psi (i)$, $R_{8,9} \leftarrow H_0$.} Here $\psi (i)$ is a multi-lane Argon2 block index defined in Eq.~\ref{eq:blockindex}. Assign $R'\leftarrow R$.
   \item Apply the modified Blake2b round
$\mathcal{P}$ first rowwise, and then columnwise to get $Z$:
\begin{align*}
    (Q_0,Q_1,\ldots,Q_7) &\leftarrow \mathcal{P}(R_0,R_1,\ldots,R_7);\\
        (Q_8,Q_9,\ldots,Q_{15})&\leftarrow \mathcal{P}(R_8,R_9,\ldots,R_{15});\\
        \ldots&\\
        (Q_{56},Q_{57},\ldots,Q_{63})&\leftarrow \mathcal{P}(R_{56},R_{57},\ldots,R_{63});\\[10pt]
        (Z_0,Z_8,Z_{16},\ldots,Z_{56})&\leftarrow \mathcal{P}(Q_0,Q_8,Q_{16},\ldots,Q_{56});\\
        (Z_1,Z_9,Z_{17},\ldots,Z_{57})&\leftarrow     \mathcal{P}(Q_1,Q_9,Q_{17},\ldots,Q_{57});\\
        \ldots&\\
        (Z_7,Z_{15},Z_{23},\ldots,Z_{63})&\leftarrow \mathcal{P}(Q_7,Q_{15},Q_{23},\ldots,Q_{63}).
  \end{align*}
  Finally,  output $Z\oplus R'$:
  $$
  F:\; (X,Y)\; \rightarrow\; R = X\oplus Y\;\xrightarrow{i,H_0}\;R' \xrightarrow{\mathcal{P}}\;Q\;\xrightarrow{\mathcal{P}}\;Z\;
  \rightarrow \;Z\oplus R'.
  $$
\end{enumerate}

The modified Argon2 algorithm uses the same indexing function $\phi$, but discards 48 (out of 1024) block bytes to reserve place for $i$ and $H_0$ as inputs to $F$. Therefore the tradeoff attack quality grows by less than 5\%.

      \subsection{MTP-Argon2}

As a concrete application, we suggest a cryptocurrency proof-of-work based on the modified Argon2d with 4 parallel lanes. The challenge $I$ should be part of the auxiliary data $U$, whereas $P$ and $S$ can be set to full-zero 16-byte strings. The $\psi$ function that maps one-dimensional indices to the two-dimensional Argon2 is given in Section~\ref{sec:ph}.

We aim to make this PoW unattractive for botnets, so we suggest using 2 GiB of RAM, which is very noticeable (and thus would likely alarm the user), while being bearable for the regular user, who consciously decided to use his desktop for mining.   Argon2 runs in 0.6 cycles per byte, so on a 3 GHz CPU a single call to 2-GiB Argon2d would take 0.4 seconds. The Merkle tree computation could take longer as we have to hash 2 GiB of data splitted into 1 KiB blocks.

 We suggest the full Blake2b for the hash function $H$ but the reduced  4-round Blake2b for the Merkle tree hash function $G$, as currently no preimage or collision attack exist on even a 3-round Blake2b (and very strong cryptography is not necessary in this part of the Prover algorithm). The 4-round Blake2b with 128-bit output thus provides 64-bit security, which is even bigger than we expected from a PoW. The reduced Blake2b would then hash 2 GiB in 0.8 seconds on a single core, so with multiple cores we expect it to run within 0.3-0.4 seconds and  the total computation time being around 0.7 seconds, as  calls to 4-round Blake2b can be interleaved with memory requests in Argon2.

\paragraph{Determining optimal $L$} Consider an ASIC-equipped adversary who employs the cheating strategies given in~Section~\ref{sec:cheat}. His advantage in the time-area product will differ for the initialization phase (when computing $T$ blocks of Argon2) and for the search phase (when he searches for $L$ blocks satisfying the difficulty test). The difference is due to the  overhead to compute the skewed blocks. The relative importance of these two phases will be determined by the ratio between the initialization time and average lifetime of a challenge~\footnote{For instance, if initialization takes 1 second, and a block appears every 20 seconds, the ratio is 20.}.  Using data from Table~\ref{tab:generic3} and interpolating 
for memory fraction parameters not given there, we can determine the adversarial  advantage by Equation~\eqref{eq:at} in both phases for fixed $L,\alpha, \epsilon,\delta$. 

Table~\ref{tab:optimal} provides minimum $L$ for given ratio ($L$-search over $T$-fill) and maximum tolerable advantage.
\begin{table}[ht]
$$
\begin{array}{||c||c|c|c|c|c|}
\hline\hline
\text{Advantage}\backslash\text{Ratio}& 0.1 & 1 & 10 &100&1000\\
\hline\hline
2 & 75&80&92&108&125\\ \hline
4 &66 &70&82&96&111\\ \hline
8 & 57&61&71&84& 97\\ \hline
16 &49 &52&61& 73&85 \\ \hline
32 &41&44&52&63&73 \\
\hline\hline
\end{array}
$$
\caption{Optimal number of openings as a function of search-initialization ratio and tolerable adversary advantage.}\label{tab:optimal}\end{table}

We  suggest $L=70$, so that the entire proof consists of about 140 blocks and their openings (we open $3L$ blocks, but $X[i_j]$ can be computed and does not need to be stored in the proof). The opening for $X[i_j]$ mostly coincides with that of $X[i_j-1]$: in $1/2$ of cases the data is the same, in $1/4$ of cases an extra 128-bit block is needed, in $1/8$ -- two extra blocks, thus 128 extra bits on average. In total we need $140\cdot 1024 + 140\cdot 21.5\cdot 16 \approx 187$ KiB.

\paragraph{Future Proofing}
In order to foresee future proofing, such as availability of cheaper RAM in the future,  MTP can be used with larger RAM sizes in the range of 2-8 GiB. Due to the very fast speed of Argon2 the initialization would be still within a few seconds and thus the PoW would be still progress-free.
Moreover MTP memory capacity can be a function  of the current difficulty and the block number.
For the same difficulty level, we expect that the 
search-initialization ratio would decrease proportionally to the increase of memory size. Table~\ref{tab:optimal} implies that $X$-fold decrease in the ratio would require $L$ bigger by $X$ elements, i.e. slightly.

 \paragraph{Comparison to other fast PoW}
 Thanks to the miner competition in Zcash, we know that the Equihash~\cite{Holiday,Zcash} can require the attacker to have 150 MB of RAM while producing 34 solutions per second on the i7-7700 CPU. An increase in the memory requirements would at least linearly increase the running time, so we conclude that a version of Equihash  which enforces an attacker to have 2 GB of RAM would not be able to fill the memory faster than in 0.5 seconds, and probably much more as the data would no longer fit into the L3 cache.
 
 Itsuku~\cite{Itsuku} is a recently proposed design by Coelho et al. It differs from MTP-Argon2 in smaller 64-byte blocks, full-blown Blake2b for the internal compression function  $F$ and two data-dependent memory accesses per block (plus two more data-independent ones). Although the benchmarks of Itsuku are not available, we could guesstimate that the 32-fold increase in the number of memory accesses would result in 10-15x decrease in performance. Therefore, to fit in 1 second Itsuku would have to be run with 100-200 MiB at most.

\subsection{MTP as a tool for time-lock puzzles and timestamping}  The paradigm of inherently sequential computation was developed by~\cite{CaiLSY93} in the application to CPU benchmarking and~\cite{rivest1996time} for timestamping, i.e. to certify that the document was generated certain amount of time in the past. Rivest et al. suggested \emph{time-lock puzzles} for this purpose. In our context, a time-lock puzzle solution is a proof-of-work that has lower bound on the running time assuming unlimited parallelism.

The verifier in~\cite{rivest1996time,JerschowM10}  selects a prime product $N=pq$ and asks the prover to compute the exponent $2^{2^D} \pmod{N}$ fpr some $D\approx N$. It is conjectured that the prover who does not know the factors can not exponentiate faster than do $D$ consecutive squarings. In turn, 
the verifier can verify the solution by computing the exponent $2^D$ modulo $\phi(N)$, which takes $\log(D)$ time. So far the conjecture has not been refuted, but the scheme inherently requires a secret held by the verifier, and thus is not suitable for proofs-of-work without secrets, as in cryptocurrencies.

Time-lock puzzles without secrets were suggested by Mahmoody et al.~\cite{MahmoodyMV13}. Their construction is a graph of hash computations, which is based on depth-robust graphs similarly to~\cite{DziembowskiFKP13}. 
The puzzle is a deterministic graph  such that removing any constant fraction of nodes keeps its depth above the constant fraction of the original one (so the parallel computation time is lower bounded). A Merkle tree is put atop of it with its root determining a small number of nodes to open. Therefore, a cheater who wants to compute the graph in less time has to subvert too many nodes and is likely to be caught.
As~\cite{DziembowskiFKP13}, the construction by Mahmoody et al., if combined with the difficulty filter, is subject to the grinding attack described above.

The MTP-Argon2 construction can be viewed as a time-lock puzzle and an improvement over these schemes.  First, the difficulty filter is explicitly based on the grinding attack, which makes it a legitimate way to solve the puzzle. Secondly, it
is much faster due to high speed of Argon2d. The time-lock property comes from the fact that the computation chain can not be parallelized as the graph structure is not known before the computation. 

Suppose that   MTP-Argon2 is parallelized by the additional factor of $R$ so that each core computes a chain of length about $T/R$. Let core $j$ compute $j$-th (out of $R$) chain, chronologically. Then by step $i$ each core has computed $i$ blocks and has not computed $T/R-i$ blocks, so  the probability that core $j$ requests a block that has not been computed is 
$$
\frac{(j-1)(T/R-i)}{(j-1)T/R+i}\leq \frac{(j-1)(T/R-i)}{jT/R}.
$$
Summing by all $i$, we obtain that core $j$ misses at least $\frac{T(1-1/j)}{2R}$,
so the total fraction of inconsistent blocks is about $0.5-\frac{\ln R}{2R}$. Therefore, $\epsilon$ quickly approaches $0.5$, which is easily detectable. We thus conclude that a parallel implementation of MTP-Argon2 is likely to fail the Merkle tree verification.

\subsection{Resistance to attacks specific for MTP-Argon2 v1.0}

Dinur and Nadler~\cite{DinurN17} and Bevand~\cite{Bevand17} demonstrated some vulnerabilities in the original design of MTP. We now explore these vulnerabilities and show that they are not present in the current version of MTP.
\begin{itemize}
    \item In the original version of MTP the challenge $I$ was not used in the internal compression function $F$. As a result, an attacker could reuse~\cite[Section 4.1]{DinurN17} most of the state blocks from another challenge claiming that they are derived from the current challenge. Furthermore, after some (moderately hard) amount of precomputation an attacker could produce~\cite[Section 4.3]{DinurN17} chunks of blocks for which the function $\phi$ outputs indices from a certain range $\mathcal{R}$, so that only blocks from $\mathcal{R}$ are stored. These chunks could be used for multiple challenges and thus could amortize the precomputation cost.
    
    Now $I$ undergoes a collision-resistant hash function to produce $H_0$, and this hash of $I$ is used in all calls to $F$. As a result, the block pairs  $(X[\phi(i_j)],X[i_j-1])$ from previous proofs would produce a different challenge-dependent  $X[i_j]$ and thus would fail the Merkle tree verification. Note that verifying the opening of $X[1]$, as suggested in ~\cite{DinurN17}, does not prevent the attack as an attacker could recompute as many first blocks as he wants and then reuse the rest.
    \item In the original version of MTP the openings for blocks $X[i_j]$ were not provided by the Prover. As a result, an attacker could take~\cite[Section 4.1]{DinurN17} arbitrary values for all $X[i]$ as $X[i_j]$ was used only to compute $Y_j$, but was not verified against the Merkle root. As now the opening for $X[i_j]$ is provided (though the block itself does not need to be stored), this attack no longer works.
    \item In the original version of MTP the block index $i$ was not used in $F$. As a result, in the case of multi-lane Argon2 an adversary could use the same blocks for all lanes in the first segment as only the current lane can be referred to by $\phi$ in the first segment (\cite[Attack 1]{Bevand17}). For the other segments, it was apparently possible to generate a chunk of size $m$ with complexity $2^{m/2}$ so that the function $\phi$ for these indices maps to the other lanes (\cite[Attack 4]{Bevand17}). Then the chunk can be concatenated in multiple copies since $\phi$ still maps to the other lanes. As each compression function computation now uses block index $i$, the attack no longer works.
    \item In the original MTP specification it was not asserted (though was implied) in the verifier's algorithm that $\phi(i_j)$ and $i_j-1$ are the positions with respect to which the opening is verified. As a result an attacker could provide identical openings for every $j$ (\cite[Attack 2]{Bevand17}). The specification now explicitly requires that opened block positions are provided in the proof and are verified.
\end{itemize}

\section{Memory-hard encryption on low-entropy keys}\label{sec:mhe}

\subsection{Motivation}

In this section\footnote{This section didn't change compared to the original USENIX Security paper.} we approach standard encryption from the memory-hardness perspective. A typical approach to hard-drive encryption is to derive the master key from the user password and  then use it to encrypt chunks of data in a certain mode of operation such as XTS~\cite{martin2010xts}. The major threat, as to other password-based security schemes, are low-entropy passwords. An attacker, who gets access to the hard drive encrypted with such password, can determine the correct key and then decrypt within short time.

A countermeasure could be to use a memory-hard function for the key derivation, so that the trial keys can be produced only on memory-rich machines.  However, the trial decryption could still be performed on special memoryless hardware given these keys. We suggest a more robust scheme which covers this type of adversaries and eventually requires that the entire attack code have permanent access to large memory. 

\subsection{Requirements}

We assume the following setting, which is inspired by typical disk-encryption applications. The data consists of multiple chunks $Q\in\mathcal{Q}$, which can be encrypted and decrypted independently.  The only secret that is available to the encryption scheme $\mathcal{E}$ is the user-input password $P\in\mathcal{P}$, which has sufficiently low entropy to be memorized (e.g., 6 lowercase symbols). The encryption syntax is then as follows:
$$
\mathcal{E}:\; \mathcal{P}\times\mathcal{S}\times \mathcal{Q} \rightarrow \mathcal{C},
$$
where $S\in\mathcal{S}$ is associated data, which may contain salt, encryption nonce or IV, chunk identifier, time, and other secondary input; and $C\in\mathcal{C}$ is ciphertext. $S$ serves both to simplify ciphertext identification (as it is public) and to ensure certain cryptographic properties. For instance, unique salt or nonce prevents repetition of ciphertexts for identical plaintexts. We note that in some settings due to storage restriction the latter requirement can be dropped. Decryption then is naturally defined and we omit its formal syntax.

In our proposal we do not restrict the chunk size. Even though it can be defined for chunks as small as disk sectors, the resistance to cracking attacks will be higher for larger chunks, up to a megabyte long.

A typical attack setting is as follows. An attacker obtains the encrypted data via some malicious channel or installs malware and then tries different passwords to decrypt it. For the sake of simplicity, we assume that the plaintext contains sufficient redundancy so that a successful guess can be identified easily. Therefore, the adversary tries $D$ passwords from his dictionary $\mathcal{D}\subset \mathcal{P}$. Let $T$ be the time needed for the fastest decryption operation that provides partial knowledge of plaintext sufficient to discard or remember the password, and $A_0$ be the chip area needed to implement this operation. Then the total amount of work performed by the adversary is
$$
W = D\cdot T\cdot A_0.
$$
At the same time, the time to encrypt $T'$ for a typical user should not be far larger than $T$. Our goal is to maximize $W$ with keeping $T'$ the same or smaller.

The memory-hard functions seem to serve perfectly for the purpose of maximizing $W$. However, it remains unclear how to combine such function $\mathcal{F}$ with $\mathcal{E}$ to get \emph{memory-hard encryption} (MHE). 

Now we formulate some additional features that should be desirable  for such a scheme:
\begin{itemize}
\item The user should be able to choose the requested memory size $A$ independently of the  chunk length $|Q|$. Whereas the chunk length can be primarily determined by the CPU cache size, desirable processing speed, or the hard drive properties, the memory size determines the scheme's resistance to cracking attacks. 
\item The memory can be allocated independently for each chunk or reused. In the former case the user can not allocate too much memory as the massive decryption would be too expensive. However, for the amounts of memory comparable to the chunk size the memory-hard decryption should take roughly as much as memoryless decryption.
If the allocated memory is reused for distinct chunks, much more memory can be allocated as the allocation time can be amortized. However, the decryption latency would be quite high. We present both options in the further text.
\item Full ciphertext must be processed to decrypt a single byte. This property clearly makes $T$ larger since the adversary would have to process an entire chunk to check the password. At the same time, for disk encryption it should be fine to decrypt in the ``all-or-nothing'' fashion, as the decryption time would still be smaller than the user could wait. 
\item Encryption  should be done in one pass over data. It might sound desirable that the decryption should be done in one pass too. However, this would contradict the previous requirement. Indeed, if the decryption can be done in one pass, then the first bytes of the plaintext can be determined without the last bytes of the ciphertext\footnote{The similar argument is made for the online authenticated ciphers in~\cite{HoangRRV15}.}. 
\item Apart from the memory parameter, the total time needed to allocate this memory should be tunable too. It might happen that the application does not have sufficient memory but does have time. In this case, the adversary can be slowed down by making several passes over the memory during its initialization (the memory-hard function that we consider support this feature).
\end{itemize}
Our next and final requirement comes from  adversary's side. When the malware is used, the incoming network connection and memory for this malware can be limited. Thus, it would be ideal for the attacker if the memory-intensive part can be delegated to large machines under attacker's control, such as botnets. If we just derived the secret-key $K$ for encryption as the output of the memory-hard hash function $\mathcal{F}$, this would be exactly this case. An adversary would then run $\mathcal{F}$ for dictionary $D$ on his own machine, produce the set $\mathcal{K}$ of keys, and supply them to malware (recall that due to low entropy there would be only a handful of these keys). Thus the final requirement should be the following:
\begin{itemize}
\item During decryption, it should be impossible to delegate the entire memory-hard computation to the external device without accessing the ciphertext. Therefore, there could be no memory-hard precomputation.
\end{itemize}

\subsection{Our scheme}

Our scheme is based on a recent proposal by Zaverucha~\cite{Zaverucha15}, who addresses similar properties in the scheme based on  Rivest's All-or-Nothing transform (ANT). However, the scheme in~\cite{Zaverucha15} does not use an external memory-hard function, which makes it memory requirements inevitably bound to the chunk size. Small chunks but large memory is impossible in~\cite{Zaverucha15}.

\begin{figure*}[ht]
  \ifpdf
\begin{center}
  \includegraphics[scale=0.9]{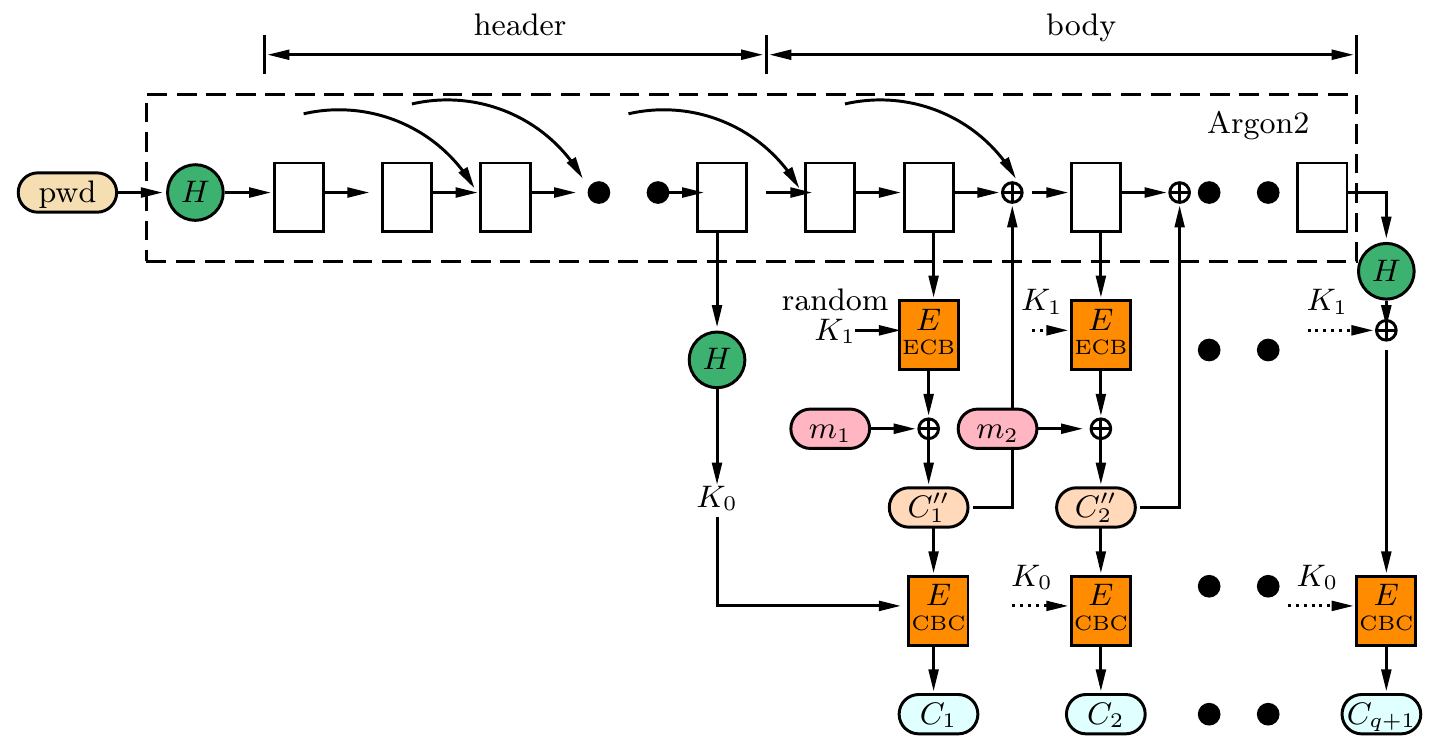}
  \caption{MHE: Disk encryption using memory-hard function Argon2. }\label{fig:mhe}
\end{center}
\fi
  \end{figure*}

Our  proposal is again based on the All-or-Nothing transformation, though we expect that similar properties can be obtained with deterministic authenticated encryption scheme as a core primitive. The chunk length $q$ (measured in blocks using by $\mathcal{F}$) and memory size $M \geq q$ are the parameters as well as some blockcipher $E$ (possibly AES). First, we outline the scheme where the memory is allocated separately for each chunk. The reader may also refer to Figure~\ref{fig:mhe}. 

\begin{algorithm}
\textbf{Input:} Password $P$, memory size $M$, associated data $S$, chunk $Q$, number of iterations $t$, memory-hard function $\mathcal{F}$ (preferably Argon2), blockcipher $E$, cryptographic hash function $H$ (e.g. SHA-3).
\begin{enumerate}
\item Run $\mathcal{F}$ on $(P,S)$ with input parameters $M$ and $t$ but fill only $M-q$ blocks (the \emph{header}) in the last iteration. Let $X_0$ be the last memory block produced by $\mathcal{F}$.
\item Produce  $K_0 = H(X_0)$ --- the first session key.
\item Generate a random session key $K_1$.
\item Generate the remaining blocks $X_1$, $X_2$, $\ldots$, $X_q$ (\emph{body}) for $\mathcal{F}$ as follows. We assume that each chunk $M$ consists of smaller blocks $m_1$, $m_2$, $\ldots$, $m_q$ of length equal to the block size of $\mathcal{F}$. For each $i\geq 1$:
\begin{itemize}
\item Encrypt $X_{i-1}$ by $E$ in the ECB mode under $K_1$ and get the intermediate ciphertext block $C_i'$.
\item Add the chunk data: $C_i'' = C_i' \oplus m_i$.
\item Encrypt $C_i''$ under $K_0$ in the CBC mode and produce the final ciphertext block $C_i$. 
\item Modify the memory: $X_{i-1}\leftarrow X_{i-1} \oplus C_i''$.
\item Generate the block $X_i$ according to the specification of $\mathcal{F}$. In Argon2, the modified $X_{i-1}$ and some another block $X[\phi(X_{i-1})]$ would be used.
\end{itemize}
\item After the entire chunk is encrypted, encrypt also the key $K_1$ :
$$
C_{t+1} = E_{K_0}(H(X_t)\oplus K_1).
$$
\end{enumerate}
\textbf{Output:} $C_1,\ldots, C_{t+1}$.
\caption{Memory-hard encryption with independent memory allocation (for each chunk).}\label{alg:mhe1}
\end{algorithm}
The underlying idea is to use both the header and the body blocks to produce the ciphertext. In tun, to recompute the body blocks both the ciphertext and the header 
must be available during trial decryption.

The version of the MHE scheme which allocates the same memory for multiple chunks is very similar. The $S$ input is ignored at the beginning, so that the header memory blocks do not depend on the data. Instead, we set $K_0 = H(X_0,S)$, so that the body blocks are affected by $S$ and $M$, and thus are different for every chunk. In this case the body blocks have to be stored 
separately and should not overwrite the header blocks for $t>1$.

Let us verify that the scheme in Algorithm~\ref{alg:mhe1} satisfies the properties we listed earlier:
\begin{itemize}
\item The allocated memory size $M$ can be chosen independently of the chunk length $q$ (as long as $M> q$).
\item The body memory blocks are allocated and processed for each chunk independently. In addition, the header blocks are also processed independently for each chunk in the single-chunk version.
\item In order to decrypt a single byte of the ciphertext, an adversary would have to obtain $K_1$, which can be done only by running $\mathcal{F}$ up to the final block, which requires all $C_i''$, which are in turn must be derived from the ciphertext blocks.
\item Encryption needs one pass over data, and decryption needs two passes over data.
\item The total time needed to allocate and fill the header is tunable.
\item The computation of the body memory blocks during decryption can not be delegated, as it requires knowledge both of the header and the ciphertext. It might be possible to generate the header on an external machine, but then random access to its blocks to decrypt the ciphertext is required.
\end{itemize}
We note that properties 1, 5, and 6 are not present in~\cite{Zaverucha15}. 

\paragraph{Security} First, we address traditional CPA security. We do not outline the full proof here, just the basic steps. We assume that the adversary does not have access to the internals of Argon2, and that blockcipher $E$ is a secure PRF. Next, we assume collision-resistance of the compression function $F$ used in $\mathcal{F}$. Given that, we prove that all the memory blocks are distinct, which yields the CPA security for $C'$. From the latter we deduce the CPA security for the final ciphertext. We note that in the case when the collision-resistance of $F$ can not be guaranteed,  we may additionally require that $X_i$ undergo hashing by a cryptographic hash function $H'$ before encryption, so that the plaintext blocks are still distinct. All these properties hold up to the birthday bound of the blockcipher.

Next, we figure out the tradeoff security. The genuine decrypting user is supposed to spend $M$ memory blocks for $\mathcal{F}$ and $q$ memory blocks to store the plaintext
and intermediate variables (if the ciphertext can be overwritten, then these $q$ blocks are not needed). Suppose that an adversary wants to use $\alpha M$ memory for header and body. Then each missing block, if asked during decryption, must be recomputed making $C(\alpha)$ calls to $F$. The best such strategy for Argon2, described in~\cite{trade-att}, yields $C(\alpha)$ that grows
exponentially in $1/\alpha$. For example, using $1/5$ of memory, an adversary would have to make 344 times as many calls to $F$, which makes a memory-reducing encryption cracking inefficient even on special hardware.

\paragraph{Performance} We suggest taking $l=4$ in Argon2 in order to fill the header faster using multiple cores, which reportedly takes 0.7 cpb (about the speed of AES-GCM and AES-XTS). The body has to be filled sequentially as the encryption process is sequential. As AES-CBC is about 1.3 cpb, and we use two of it, the body phase should run at about 4 cpb. In a concrete setting,
suppose that we tolerate 0.1 second decryption time (about 300 Mcycles) for the 1-MiB chunk. Then we can take the header as large as 256 MiB, as it would be processed in 170 Mcycles + 4 Mcycles for the body phase.

\section{Conclusion}

We have introduced the new paradigm of egalitarian computing, which suggests amalgamating arbitrary computation with a memory-hard function to enhance the security against off-line adversaries equipped with powerful tools (in particular with optimized hardware). We have reviewed password hashing and proofs of work as applications where such schemes are already in use or are planned to be used. We then introduce two more schemes in this framework. The first one is MTP, the progress-free proof-of-work scheme with fast verification based on the memory-hard function Argon2, the winner of the Password Hashing Competition. The second scheme pioneers the memory-hard encryption --- the security enhancement for password-based disk encryption, also based on Argon2.

\section*{Acknowledgement}
We would like to thank Itai Dinur, Niv Nadler, Mark Bevand, Fabien Coelho (and his co-authors) for sharing their results with us and for discussions. We also thank Zcoin project~\cite{Zcoin} for organizing the MTP bounty challenge and for partial funding of this work.

{\footnotesize \bibliographystyle{acm}
\bibliography{xorhard}}

\begin{thebibliography}{10}

\bibitem{bitasic}
Avalon asic's 40nm chip to bring hashing boost for less power, 2014.
\newblock \url{
  http://www.coindesk.com/avalon-asics-40nm-//chip-bring-hashing-boost-less-power/}.

\bibitem{bitcoinhard}
Bitcoin: Mining hardware comparison, 2014.
\newblock available at
  \url{https://en.bitcoin.it/wiki/Mining_hardware_comparison}.

\bibitem{comp}
{Password Hashing Competition}, 2015.
\newblock \url{https://password-hashing.net/}.

\bibitem{MillerCohen16}
2016.
\newblock Andrew Miller, Bram Cohen, private communication.

\bibitem{Zcoin}
Zcoin, 2017.
\newblock \url{http://zcoin.io/}.

\bibitem{AbadiBW03}
{\sc Abadi, M., Burrows, M., and Wobber, T.}
\newblock Moderately hard, memory-bound functions.
\newblock In {\em {NDSS}'03\/} (2003), The Internet Society.

\bibitem{Andersen14}
{\sc Andersen, D.}
\newblock A public review of cuckoo cycle.
\newblock \url{http://www.cs.cmu.edu/~dga/crypto/cuckoo/analysis.pdf}, 2014.

\bibitem{back2002hashcash}
{\sc Back, A.}
\newblock Hashcash -- a denial of service counter-measure, 2002.
\newblock available at \url{http://www.hashcash.org/papers/hashcash.pdf}.

\bibitem{BernsteinL13}
{\sc Bernstein, D.~J., and Lange, T.}
\newblock Non-uniform cracks in the concrete: The power of free precomputation.
\newblock In {\em ASIACRYPT'13\/} (2013), vol.~8270 of {\em Lecture Notes in
  Computer Science}, Springer, pp.~321--340.

\bibitem{Bevand17}
{\sc Bevand, M.}
\newblock Attacks on merkle tree proof.
\newblock Blog of M. Bevand, 2017.
\newblock \url{http://blog.zorinaq.com/attacks-on-mtp/}.

\bibitem{trade-att}
{\sc Biryukov, A., and Khovratovich, D.}
\newblock Tradeoff cryptanalysis of memory-hard functions.
\newblock In {\em Asiacrypt'15\/} (2015).
\newblock available at \url{http://eprint.iacr.org/2015/227}.

\bibitem{Argon2}
{\sc Biryukov, A., and Khovratovich, D.}
\newblock Argon2: new generation of memory-hard functions for password hashing
  and other applications.
\newblock In {\em Euro S\&P'16\/} (2016).
\newblock available at \url{https://www.cryptolux.org/images/0/0d/Argon2.pdf}.

\bibitem{Holiday}
{\sc Biryukov, A., and Khovratovich, D.}
\newblock Equihash: Asymmetric proof-of-work based on the generalized birthday
  problem.
\newblock In {\em NDSS'16\/} (2016).
\newblock available at \url{https://eprint.iacr.org/2015/946.pdf}.

\bibitem{CaiLSY93}
{\sc Cai, J., Lipton, R.~J., Sedgewick, R., and Yao, A.~C.}
\newblock Towards uncheatable benchmarks.
\newblock In {\em Structure in Complexity Theory Conference\/} (1993), {IEEE}
  Computer Society, pp.~2--11.

\bibitem{Itsuku}
{\sc Coelho, F., Larroche, A., and Colin, B.}
\newblock Itsuku: a memory-hardened proof-of-work scheme.
\newblock Cryptology ePrint Archive, Report 2017/1168, 2017.
\newblock \url{https://eprint.iacr.org/2017/1168}.

\bibitem{DinurN17}
{\sc Dinur, I., and Nadler, N.}
\newblock Time-memory tradeoff attacks on the {MTP} proof-of-work scheme.
\newblock In {\em {CRYPTO} {(2)}\/} (2017), vol.~10402 of {\em Lecture Notes in
  Computer Science}, Springer, pp.~375--403.

\bibitem{DworkGN03}
{\sc Dwork, C., Goldberg, A., and Naor, M.}
\newblock On memory-bound functions for fighting spam.
\newblock In {\em CRYPTO'03\/} (2003), vol.~2729 of {\em Lecture Notes in
  Computer Science}, Springer, pp.~426--444.

\bibitem{DworkN92}
{\sc Dwork, C., and Naor, M.}
\newblock Pricing via processing or combatting junk mail.
\newblock In {\em CRYPTO'92\/} (1992), vol.~740 of {\em Lecture Notes in
  Computer Science}, Springer, pp.~139--147.

\bibitem{DworkNW05}
{\sc Dwork, C., Naor, M., and Wee, H.}
\newblock Pebbling and proofs of work.
\newblock In {\em {CRYPTO}'05\/} (2005), vol.~3621 of {\em Lecture Notes in
  Computer Science}, Springer, pp.~37--54.

\bibitem{DziembowskiFKP13}
{\sc Dziembowski, S., Faust, S., Kolmogorov, V., and Pietrzak, K.}
\newblock Proofs of space.
\newblock In {\em CRYPTO'15\/} (2015), R.~Gennaro and M.~Robshaw, Eds.,
  vol.~9216 of {\em Lecture Notes in Computer Science}, Springer, pp.~585--605.

\bibitem{giridhar2013dram}
{\sc Giridhar, B., Cieslak, M., Duggal, D., Dreslinski, R.~G., Chen, H., Patti,
  R., Hold, B., Chakrabarti, C., Mudge, T.~N., and Blaauw, D.}
\newblock Exploring {DRAM} organizations for energy-efficient and resilient
  exascale memories.
\newblock In {\em International Conference for High Performance Computing,
  Networking, Storage and Analysis 2013\/} (2013), ACM, pp.~23--35.

\bibitem{HoangRRV15}
{\sc Hoang, V.~T., Reyhanitabar, R., Rogaway, P., and Viz{\'{a}}r, D.}
\newblock Online authenticated-encryption and its nonce-reuse
  misuse-resistance.
\newblock In {\em {CRYPTO}'15\/} (2015), R.~Gennaro and M.~Robshaw, Eds.,
  vol.~9215 of {\em Lecture Notes in Computer Science}, Springer, pp.~493--517.

\bibitem{HopcroftPV77}
{\sc Hopcroft, J.~E., Paul, W.~J., and Valiant, L.~G.}
\newblock On time versus space.
\newblock {\em J. ACM 24}, 2 (1977), 332--337.

\bibitem{Zcash}
{\sc Hopwood, D., Bowe, S., Hornby, T., and Wilcox, N.}
\newblock Zcash protocol specification. version 2017.0-beta-2.7, 2017.
\newblock \url{https://github.com/zcash/zips/raw/master/protocol/protocol.pdf}.

\bibitem{JerschowM10}
{\sc Jerschow, Y.~I., and Mauve, M.}
\newblock Offline submission with {RSA} time-lock puzzles.
\newblock In {\em {CIT}\/} (2010), {IEEE} Computer Society, pp.~1058--1064.

\bibitem{momentum}
{\sc Lorimer, D.}
\newblock Momentum -- a memory-hard proof-of-work via finding birthday
  collisions, 2014.
\newblock available at \url{http://www.hashcash.org/papers/momentum.pdf}.

\bibitem{MahmoodyMV13}
{\sc Mahmoody, M., Moran, T., and Vadhan, S.~P.}
\newblock Publicly verifiable proofs of sequential work.
\newblock In {\em {ITCS}\/} (2013), {ACM}, pp.~373--388.

\bibitem{fpga}
{\sc Malvoni, K.}
\newblock Energy-efficient bcrypt cracking, 2014.
\newblock Passwords'14 conference.

\bibitem{martin2010xts}
{\sc Martin, L.}
\newblock Xts: A mode of aes for encrypting hard disks.
\newblock {\em IEEE Security \& Privacy}, 3 (2010), 68--69.

\bibitem{EthashAnalysis}
{\sc Miller, A., Warner, B., and Wilcox, N.}
\newblock Analysis of ethash, ethereum's proof-of-work puzzle, 2015.
\newblock
  \url{https://github.com/LeastAuthority/ethereum-analyses/blob/master/PoW.md}.

\bibitem{nakamoto09}
{\sc Nakamoto, S.}
\newblock Bitcoin: A peer-to-peer electronic cash system.
\newblock \url{http://www. bitcoin.org/bitcoin.pdf}.

\bibitem{ParkPAFG15}
{\sc Park, S., Pietrzak, K., Alwen, J., Fuchsbauer, G., and Gazi, P.}
\newblock Spacecoin: {A} cryptocurrency based on proofs of space.
\newblock {\em {IACR} Cryptology ePrint Archive 2015\/} (2015), 528.

\bibitem{percival2009stronger}
{\sc Percival, C.}
\newblock Stronger key derivation via sequential memory-hard functions.
\newblock \url{http://www.tarsnap.com/scrypt/scrypt.pdf}.

\bibitem{Pippenger77}
{\sc Pippenger, N.}
\newblock Superconcentrators.
\newblock {\em {SIAM} J. Comput. 6}, 2 (1977), 298--304.

\bibitem{rivest1996time}
{\sc Rivest, R.~L., Shamir, A., and Wagner, D.~A.}
\newblock Time-lock puzzles and timed-release crypto.
\newblock \url{https://people.csail.mit.edu/rivest/pubs/RSW96.pdf}.

\bibitem{SprengerB12}
{\sc Sprengers, M., and Batina, L.}
\newblock Speeding up {GPU-based} password cracking.
\newblock In {\em SHARCS'12\/} (2012).
\newblock available at \url{http://2012.sharcs.org/record.pdf}.

\bibitem{Thompson79}
{\sc Thompson, C.~D.}
\newblock Area-time complexity for {VLSI}.
\newblock In {\em STOC'79\/} (1979), ACM, pp.~81--88.

\bibitem{Tromp14}
{\sc Tromp, J.}
\newblock Cuckoo cycle: a memory bound graph-theoretic proof-of-work.
\newblock Cryptology ePrint Archive, Report 2014/059, 2014.
\newblock available at \url{http://eprint.iacr.org/2014/059}, project webpage
  \url{https://github.com/tromp/cuckoo}.

\bibitem{OorschotW99}
{\sc van Oorschot, P.~C., and Wiener, M.~J.}
\newblock Parallel collision search with cryptanalytic applications.
\newblock {\em J. Cryptology 12}, 1 (1999), 1--28.

\bibitem{Wagner02}
{\sc Wagner, D.}
\newblock A generalized birthday problem.
\newblock In {\em CRYPTO'02\/} (2002), vol.~2442 of {\em Lecture Notes in
  Computer Science}, Springer, pp.~288--303.

\bibitem{EthereumYellow}
{\sc Wood, G.}
\newblock Ethereum: a secure decentralised generalised transaction ledger,
  2017.
\newblock \url{http://gavwood.com/paper.pdf}.

\bibitem{Zaverucha15}
{\sc Zaverucha, G.}
\newblock Stronger password-based encryption using all-or-nothing transforms.
\newblock available at \url{http://research.microsoft.com/pubs/252097/pbe.pdf}.

\end{thebibliography}
\appendix

\section{Merkle hash trees}\label{sec:tree}
We use Merkle hash trees  in the following form. A prover $P$ commits to $T$  blocks $X[1], X[2], \ldots, X[T]$ by computing  the hash tree where the blocks $X[i]$ are at leaves at depth $\log T$ and nodes compute hashes of their branches. For instance, for 
$T=4$ and hash function $G$  prover $P$ computes and publishes
$$
\Phi = G(G(X[1],X[2]), G(X[3],X[4])).
$$
Prover stores all blocks and all intermediate hashes. In order to prove that he knows, say, $X[5]$ for $T=8$, (or to \emph{open} it) he discloses the hashes needed to reconstruct the path from $X[5]$ to $\Phi$:
\begin{multline*}
\mathrm{open}(X[5]) = (X[5], X[6], g_{78} = G(X[7],X[8]), \\g_{1234} = G(G(X[1],X[2]), G(X[3],X[4])),\Phi),
\end{multline*}
so that the verifier can make all the computations. If $G$ is collision-resistant, it is hard to open any block in more than one possible way.

\section{Difference to the original MTP}
\label{sec:AppB}
The version 1.2 of MTP-Argon2 presented in this paper differs from the original version 1.0 in the following:
\begin{itemize}
    \item The Argon2 compression function is modified where 3 16-byte blocks of its intermediate block $R$ are replaced with the block index $i$ and input hash $H_0$.
    \item The Merkle tree opening for $X[i_j]$ is now included, though the block itself doesn't need to be included, since it is computed from the blocks $(X[\phi(i_j)],X[i_j-1])$. Opening paths of $X[i_j-1]$  and $X[i_j]$ share most of the nodes, which can be used by efficient implementation;
    \item The positions of opened blocks are now included in the proof and  are verified;
    \item 4-round Blake2 is used in the Merkle tree generation;
    \item New "skewed blocks" attack strategy is presented in Sect.~\ref{sec:cheat}. However it does not effect the security parameter recommendations for MTP-Argon2, while it might effect other MTP-based PoWs.
\end{itemize}

\section{Memoryless computation of Itsuku}

Itsuku~\cite{Itsuku} is a new proof-of-work (PoW) proposal by Coelho, Larroche, and Colin. The authors make an attempt to fix vulnerabilities in the original version of MTP. 
This PoW, shortly, works as follows (using the MTP notation):
\begin{itemize}
    \item Independently fill $p$ lanes of memory, $T$ blocks total (64 bytes per block):
    $$
    X[i] = F(X[\phi_1(i)],X[\phi_2(i)],X[\phi_3(i)],X[\phi_4(i)]),
    $$
    where the $\phi$ functions always point to the computed blocks of the same lane. There $\phi_1(i)=i-1,\phi_4(i) = 7i/8$ (within the lane) and $\phi_2,\phi_3$ are data-dependent;
    \item Compute the Merkle tree over $X$;
    \item Using Merkle tree root $\Phi$ and nonce $N$, compute $L=9$ indices and open such blocks if the difficulty condition is met.
\end{itemize}
The compression function $F$ is an iteration of full regular Blake2.

\paragraph{Computing Itsuku with little memory} Our ``skewed blocks'' cheating strategy can be applied to Itsuku.
However, the process is different for the initialization ($T$-fill) and iteration ($L$-search) phases.

In the initialization phase the attacker stores no blocks in memory so $\alpha=0$. He also makes every $1/\epsilon$-th block inconsistent,  that he can look up blocks by $i/8$ steps  back at cost $2/\epsilon^2$. Finally, he ensures that  all the other $1/\epsilon-1$ blocks in a consistent chunk point to inconsistent block of this or other chunks (thus $\epsilon +\delta=1$) by functions $\phi_2,\phi_3$. 
In each chunk the attacker has to try
$$
\epsilon^{2-\frac{2}{\epsilon}}
$$
inconsistent blocks to satisfy this condition. Each block needs extra $1/\epsilon$ calls to $F$ to compute the reference block by $\phi_4$, so the amortized overhead is $\epsilon^{2-\frac{2}{\epsilon}}$.

In the search phase the attacker no longer needs to try many inconsistent blocks, but he has to look up the blocks referenced by $\phi_4$. As every $1/\epsilon$-th block is inconsistent, the maximum depth of the lookup recomputation tree is $1/\epsilon$ and the recomputation penalty is $2/\epsilon^2$.

Finally, an attacker has to compute the opening for the Merkle tree he does not store. However, he simply recomputes it from the beginning. Since he already found good inconsistent blocks, this step has negligible cost compared to the initialization phase.

The attacker will not be caught with probability $(1-\epsilon)^9$ (as $L=9$). Thus the total computational overhead per block for computing the $T$ blocks of memory and not being caught is
$$
\frac{\epsilon^{2-\frac{2}{\epsilon}}}{(1-\epsilon)^9},
$$
which reaches minimum at $\epsilon=0.43$ and equals $1475$. Therefore, using 1475 extra cores and virtually no memory, we can compute the initialization phase of Itsuku PoW. 

The computational overhead for the search phase is only $2/\epsilon^2 = 10$. Thus for a reasonable challenge-lifetime/initialization ratios of 10 or 100 the total number of cores for the attacker  is about 100 or less. A more careful  analysis with an analogue of Table~\ref{tab:generic3} would bring more accurate estimates.
\end{document}